\documentclass[preprint,pra,aps,showpacs,preprintnumbers,amsmath,amssymb,eqsecnum]{revtex4}

\usepackage{graphicx}
\usepackage{dcolumn}
\usepackage{bm}
\usepackage{verbatim}
\usepackage{epsfig}
\usepackage{epstopdf}

\begin{document}


\title{Analytic Examples, Measurement Models and Classical Limit of Quantum Backflow}

\author{J.M.Yearsley}
\email{jmy27@cam.ac.uk}
\affiliation{Centre for Quantum Information and Foundations, DAMTP, Centre for Mathematical Sciences, University of Cambridge, Wilberforce Road, Cambridge CB3 0WA, UK}

\author{J.J.Halliwell}%
\email{j.halliwell@imperial.ac.uk}

\author{R.Hartshorn}

\author{A.Whitby}

\affiliation{Blackett Laboratory \\ Imperial College \\ London SW7
2BZ \\ UK }



\begin{abstract}
We investigate the backflow effect in elementary quantum mechanics -- the phenomenon
in which a state consisting entirely of positive momenta may have negative current
and the probability flows in the opposite direction to the momentum.
We compute the current and flux for states consisting of superpositions of gaussian wave packets. These are experimentally realizable but the amount of backflow is small.
Inspired by the numerical results of Penz et al (M.Penz, G.Gr\"ubl, S.Kreidl and P.Wagner,
J.Phys. A39, 423 (2006)),
we find two non-trivial wave functions whose current at any time may be computed analytically and which have periods of significant backflow, in one case with a backwards flux equal to about 70 percent of the maximum possible backflow, a dimensionless number $c_{bm} \approx 0.04 $, discovered by Bracken and Melloy (A.J.Bracken and G.F.Melloy, J.Phys. A27, 2197
(1994)). This number has the unusual property of being independent of $\hbar$ (and also of all other parameters of the model), despite corresponding to an obviously quantum-mechanical effect, and we shed some light on this surprising property by considering the classical limit of backflow.
We discuss some specific measurement models in which backflow may be identified in certain measurable probabilities.

\end{abstract}



\maketitle

\newcommand\beq{\begin{equation}}
\newcommand\eeq{\end{equation}}
\newcommand\bea{\begin{eqnarray}}
\newcommand\eea{\end{eqnarray}}

\def\A{{\cal A}}
\def\D{\Delta}
\def\H{{\cal H}}
\def\E{{\cal E}}
\def\p{\partial}
\def\la{\langle}
\def\ra{\rangle}
\def\ria{\rightarrow}
\def\x{{\bf x}}
\def\y{{\bf y}}
\def\k{{\bf k}}
\def\q{{\bf q}}
\def\p{{\bf p}}
\def\P{{\bf P}}
\def\r{{\bf r}}
\def\s{{\sigma}}
\def\a{\alpha}
\def\b{\beta}
\def\e{\epsilon}
\def\U{\Upsilon}
\def\G{\Gamma}
\def\om{{\omega}}
\def\Tr{{\rm Tr}}
\def\ih{{ \frac {i} { \hbar} }}
\def\trho{{\rho}}

\def\au{{\underline \alpha}}
\def\bu{{\underline \beta}}
\def\pp{{\prime\prime}}
\def\id{{1 \!\! 1 }}
\def\half{\frac {1} {2}}

\def\jjh{j.halliwell@ic.ac.uk}

\section{Introduction}


A striking but seemingly little-known phenomenon in elementary quantum mechanics
is the backflow effect. This is the fact that, for a free particle described by a wave function centred in $x<0$ consisting entirely of positive momenta, the probability of remaining
in $x<0$ may, for certain states, {\it increase} with time. That is, the quantum-mechanical current at the origin can be negative and the probability flows ``backwards''.

This surprising and clearly non-classical effect, so far unchecked experimentally, was first noted by Allcock in his seminal work on arrival time in quantum mechanics \cite{All} and subsequently highlighted by Bracken
and Melloy in 1994, who elucidated some of its features \cite{BrMe,BrMe2,BrMe3}. In particular,
they showed that there is a limit on the total amount of backflow. Although backflow means that the probability for remaining in $x<0$ may increase with time, the increase
can be no greater than an amount $c_{bm}$, a dimensionless number computed numerically by Bracken and Melloy to be approximately $ 0.04$. Furthermore, although this effect is clearly non-classical, the number $c_{bm}$ is independent of $\hbar$ (and also of the particle mass $m$ and the time duration of backflow). For this reason Bracken and Melloy declared $c_{bm}$ to be a ``new quantum number''.

This remarkable effect has been further investigated by a number of authors.
Eveson et al significantly refined the numerical computation of $c_{bm}$ \cite{Eveson}.
Similar numerical computations were carried out by Penz et al \cite{Penz} who gave numerically obtained plots for the form of the state of maximum backflow
and also gave a rigorous account of the optimization problem involved. Muga
et al gave some analytic examples of backflow states and explored some
aspects relating to detection \cite{Muga}. Berry \cite{Ber} has also explored various aspects of backflow and related it to the phenomenon of superoscillations \cite{super}. Bracken and Melloy have also explored the effect in the Dirac equation \cite{BrMe2} and for a non-relativistic particle with constant force \cite{BrMe3}.
Furthermore, the existence of the effect is frequently noted in connection with the arrival time problem
\cite{MaLe,MBM,GRT,DeMu,Del,time,HaYe1,HaYe2,YDHH}.
The purpose of this paper is to explore and illustrate various aspects of backflow and in particular to provide concrete analytic examples of it.

We begin in Section 2 with a detailed formulation of the problem. We define the current and the flux and consider the properties of the spectrum of the flux operator, in terms of which the backflow problem is most clearly defined.

In Section 3 we give some simple examples of states with backflow using superpositions of gaussian states. The backflow for such states, is however, rather small.

In Section 4, we review and repeat the numerical computation of the maximal backflow state and eigenvalue.
We give two non-trivial wave functions with backflow which appear to match very closely the numerical solutions for the maximal backflow state by Penz et al \cite{Penz}.
The current at arbitrary times of these wave functions is computed analytically and we find that one has a backflow
of approximately 70 percent of the maximal value. This is a much larger backflow than any analytically tractable states previously discovered.

In Section 5, we consider the naive classical limit $\hbar \rightarrow 0$ of backflow, and in particular, we address the fact that the bound on backflow $c_{bm}$ discovered by Bracken and Melloy appears to be independent of $\hbar$. We show that the expected dependency on $\hbar$ reappears in realistic measurement models, where measurements are described not by exact projectors but by quasiprojectors involving parameters characterizing the imprecision of real measurements. Under these conditions the naive classical limit is restored.

In Section 6, we consider some simple measurement models and discuss the ways in which backflow may be seen in the probabilities for measurements.


We summarize and conclude in Section 7.

\section{Detailed Formulation of the Problem}

\subsection{The Flux}

We consider a free particle with initial wave function $ \psi (x)$ centred in $x<0$
and consisting entirely of positive momenta. Note that it may not of course
be perfectly localized in $x<0$, since this incompatible with positive momenta.
We consider the amount of probability flux
$F(t_1,t_2)$ crossing the origin during the time interval $[t_1,t_2]$, defined by
\bea
F(t_1, t_2 ) &=& \int_{-\infty}^0 dx \left| \psi (x,t_1) \right|^2 -  \int_{-\infty}^0 dx \left| \psi (x,t_2) \right|^2
\\
&=& \int_{t_1}^{t_2} dt \ J(t)
\label{flux}
\eea
where $J(t)$ is the usual quantum-mechanical current at the origin
\beq
J(t) = - \frac{i \hbar}{2m}\left(\psi^{*}(0,t)\frac{\partial
\psi(0,t)}{\partial x}-\frac{\partial \psi^{*}(0,t)}{\partial
x} \psi(0,t)\right)\label{cur}
\eeq
The flux is also easily rewritten in terms of the Wigner function \cite{Wig} at time t, $ W_t(p,q)$,
\beq
F(t_1, t_2) = \int_{t_1}^{t_2} dt \int dp dq \ \frac {p} {m} \delta (q) W_t(p,q)
\label{Wig}
\eeq
(For a useful review of the properties of the current and related phase space
distribution functions, see Ref.\cite{Wig2}).
It is also useful to write these expressions in an operator form. We introduce a projection operator onto the positive $x$-axis, $P = \theta (\hat x)$, and its complement, $ \bar P = 1 - P = \theta (- \hat x)$. The flux may then be written in terms of a flux operator $\hat F (t_1, t_2) $
defined by
\bea
\hat F(t_1,t_2) &=& P(t_2) - P(t_1)
\nonumber
\\
&=&  \int_{t_1}^{t_2} dt \ \dot P(t)
\nonumber \\
&=& \int_{t_1}^{t_2} dt \ \frac {i} {\hbar} [H, \theta ( \hat x) ]
\nonumber \\
&=& \int_{t_1}^{t_2} dt \ \hat J(t)
\label{fluxop}
\eea
where we have introduced the current operator
\beq
\hat J = \frac {1} {2 m} \left( \hat p \delta (\hat x) + \delta (\hat x) \hat p \right)
\label{curop}
\eeq
So Eqs.(\ref{flux}), (\ref{cur}) may also be written
\bea
F(t_1, t_2) &=& \langle \hat F(t_1, t_2) \rangle
\nonumber \\
&=& \langle \bar P(t_1) \rangle - \langle \bar P(t_2) \rangle
\nonumber \\
&=& \langle P(t_2) \rangle - \langle P(t_1) \rangle
\nonumber \\
&=& \int_{t_1}^{t_2} dt \ \langle \psi | \hat J(t) | \psi \rangle
\label{flux2}
\eea
where $J(t) =  \langle \psi | \hat J(t) | \psi \rangle$.

The flux Eq.(\ref{flux}) is a difference between two probabilities and is clearly positive when those probabilities behave according to classical intuition, i.e., when the probability of remaining
in $x<0$ decreases monotonically.
For this reason,
the flux is often proposed as the provisional semiclassical answer to the arrival time problem: what is the probability for crossing the origin during the time interval $[t_1,t_2]$? This is discussed
at length elsewhere \cite{MaLe,MBM,GRT,DeMu,Del,time,HaYe1,HaYe2}
and, although the arrival time problem
forms the backdrop to the current work, the properties of the flux pose an interesting
set of problems in themselves and it is these problems we focus on.


As indicated already, in the full quantum-mechanical case, the flux can be negative. The above formulae give some clues as to why this is the case. First of all,
since the Wigner function can be negative \cite{Wig}, Eq.(\ref{Wig}) suggests that the flux can be negative for certain states. More precisely, negative Wigner function
is a necessary condition for negative flux, which indicates that it relates to states
with interferences in position or momentum. Negative Wigner function is not a sufficient
condition since the integral in Eq.(\ref{Wig}) may yield a positive expression, even
for negative $W$.

The second clue to the possible negativity comes from the current operator Eq.(\ref{curop}): the two operators $\hat p$ and $\delta (\hat x)$ are non-negative (on states with
positive momentum), but since they do not commute, the current operator $\hat J$ is not a positive operator.

\subsection{Most Negative Flux as an Eigenvalue Problem}

Following Bracken and Melloy \cite{BrMe}, a useful way to find the states with negative flux is to look at the spectrum of the flux operator Eq.(\ref{fluxop}) (restricted to positive momenta). We thus look for the solution to the eigenvalue problem
\beq
\theta ( \hat p) \hat F(t_1, t_2) | \Phi \rangle = \lambda | \Phi \rangle
\label{evalue}
\eeq
where the states $| \Phi \rangle $ consist only of positive momenta.
(We choose an opposite sign convention to Bracken and Melloy which means that the backflow states have $\lambda < 0 $). The most negative value of the flux $F(t_1,t_2)$ is then the
most negative eigenvalue of the flux operator.

It is convenient to choose the time interval $[t_1,t_2]$ to be $[-T/2,T/2]$, as is
easily achieved by time evolving the state, and the eigenvalue equation in momentum space then reads
\beq
\frac {1} {\pi} \int_0^\infty dk \ \frac { \sin [(p^2 - k^2) T/ 4 m \hbar ] } {(p-k)} \
\Phi (k)
 = \lambda \Phi (p)
\eeq
We then define rescaled variables $u$ and $v$ by $ p = 2 \sqrt{ m \hbar / T} u$
and $ k = 2 \sqrt{ m \hbar / T} v$ and the eigenvalue equation is then
\beq
\frac {1} {\pi} \int_0^\infty dv \ \frac { \sin (u^2 - v^2)  } {(u-v)} \
\phi (v)
 = \lambda \phi (u)
\label{evalue2}
\eeq
where $ \phi (u) = (m \hbar / 4 T)^{1/4} \Phi (p)$ and is dimensionless. Note that all physical constants $\hbar, m, T$ have dropped out of this equation so that the eigenvalues $\lambda$ are dimensionless
and independent of $\hbar, m$ and $T$.
It is useful to record the result that the flux for any state $\phi (u)$ in these variables is given by
\beq
F(-T/2,T/2) =  \frac {1} {\pi}\int_0^\infty du  \int_0^\infty dv \ \phi^*(u) \ \frac { \sin (u^2 - v^2)  } {(u-v)} \phi (v)
\label{flux3}
\eeq

The eigenvalue equation Eq.(\ref{evalue2}) clearly has approximate solutions with eigenvalues close to $1$ or $0$ consisting of wave packets which cross the origin either well inside or well outside the interval $[-T/2,T/2]$. Further study of this eigenvalue equation has been carried out by a number of authors, both numerically and analytically
\cite{BrMe,Eveson,Penz}. The eigenvalue equation is real, so one may take the eigenstates $\phi(u)$ to be real valued-functions. This has the consequence that the corresponding wave function in
configuration space at time $t$, $\psi (x,t)$ has the symmetry
\beq
\psi^* (x,t) = \psi ( -x, -t)
\label{wavefn}
\eeq
as is indeed observed in the numerical solutions.
The eigenvalues
lie in the range
\beq
 - c_{bm} \le \lambda \le 1
\label{range}
\eeq
where $c_{bm}$ was computed numerically and found to be
\beq
c_{bm} \approx 0.038452
\label{cbm}
\eeq
It was conjectured in Ref.\cite{BrMe} that the spectrum is discrete in the interval $[-c_{bm},0]$ but continuous in the interval $[0,1]$. The extremizing state was given numerically by Penz et al \cite{Penz} who gave numerical evidence to suggest that its asymptotic form for large $u$ is close to $ \sin u^2 / u$, which indicates that the extremizing state is square-integrable.
We will find good analytic expressions approximating the numerical results for all $u$ in what follows.

The eigenvalue Eq.(\ref{evalue2}) was solved without any conditions on $\phi(u)$ at $u=0$
\cite{BrMe,Eveson,Penz}, as is reasonable for an integral equation.
Since $\phi(u) = 0$ for $u<0$ this means that the state could be discontinuous in momentum space and as a consequence the position width $(\Delta x)^2$ is infinite. The above asymptotic form for $\phi(u)$ in momentum space also means that $(\Delta p)^2$ is infinite.
This means that the state is somewhat unusual although there is no obvious reason to require that the widths in position or momentum space should be finite and indeed such a restriction may limit the amount of backflow. We offer no simple explanations as to why these properties hold, although there are some hints in Ref.\cite{Penz}.

At present there is no analytic account of the properties of the results Eqs.(\ref{range}), (\ref{cbm}).
Physically, backflow is related to the fact that specifying both positive momenta {\it and} asking for the probability of remaining in $x>0$ correspond to incompatible measurements in quantum mechanics. That is, it is related to the fact that the operators $\theta (\hat p)$ and $\theta (-\hat x) $ do not commute. This leads to the question, is there an analytic calculation of the most negative eigenvalue $- c_{bm}$, perhaps involving the non-commutativity of $\theta (\hat p)$ and $\theta (-\hat x) $?
We do not have an answer to this question but it remains an important question for future study.
Also, the fact that the eigenvalues are independent of $\hbar$ means there are some potential problems with the naive classical limit and we address this below.

Note that the eigenvalues are independent of $T$. This simply means that the duration of a period of negative current can be arbitrarily long, as long as the total flux over that time period is bounded from below by $-c_{bm}$, that is,
\beq
\int_{-T/2}^{T/2} dt \ J(t) \ge - c_{bm}
\eeq
This means that a relationship of the form
\beq
T \ J (\xi) \ge - c_{bm}
\eeq
holds, for some time $\xi$ in the interval $[-T/2,T/2]$. These relations also imply that
there is no restriction on the current being arbitrarily negative, as long as it is negative for a sufficiently short time.


\section{Backflow for superpositions of gaussians}\label{secgauss}

Bracken and Melloy gave two explicit examples of states displaying backflow \cite{BrMe}.  Although these examples served to demonstrate the existence of the effect, the particular states chosen were rather unphysical. Muga et al gave an example of a backflow state consisting of a gaussian in momentum space but restricted to $ p>0$ \cite{Muga}, which means that the wave function in configuration space is not a simple function. Berry also gave some simple examples \cite{Ber}, essentially plane waves, similar to those considered in Ref.\cite{BrMe}

In this section we show that the backflow effect also arises in the more familiar, and also potentially experimentally realisable, setting of a superposition of two gaussian wavepackets.
(This example does not seem to have been considered previously, other than the closely related result in Ref.\cite{Muga}). A single gaussian has positive Wigner function so must have positive current, Eq.(\ref{Wig}). The Wigner function of a superposition of Gaussians may, however, be negative in some regions, so the current can be negative.

Of course the problem with using gaussian wavepackets is that they have support on both positive and negative momentum, and so as well as demonstrating the appearance of negative current, we will also have to show that this is not the result of any initially negative momentum. We will see that whilst superpositions of gaussian states do indeed give rise to backflow, the size of the effect is considerably smaller than the theoretical maximum.

We begin with the simple case of a superposition of two plane waves, as considered in Ref.\cite{BrMe}. This can be turned into a more physical state by replacing the plane waves with gaussians tightly peaked in momentum, without affecting the basic conclusion that the state displays backflow for well chosen values of the various parameters. Normalizable states are necessary in order to have a properly normalized flux. In this Section we work in units in which $\hbar = 1$
and we set the particle mass $m=1$.

We start with the unnormalised state,
\beq
\psi'(x,t)=\sum_{k=1,2}A_{k}\exp[i p_{k}(x-p_{k}t)]\label{jsec1}
\eeq
where the $A_{k}$ are taken to be real.
(One could of course add to each component an arbitrary phase, but this is an unneccesssary complication.) The current at the origin for this state is given by,
\beq
J(t)=A_{1}^{2}p_{1}+A_{2}^{2}p_{2}+A_{1}A_{2}(p_{1}+p_{2})\cos[(E_{1}-E_{2})t]\label{jsec2}
\eeq
This oscillates between a maximum value of $(A_{1}p_{1}+A_{2}p_{2})(A_{1}+A_{2})$ and a minimum value of $(A_{1}p_{1}-A_{2}p_{2})(A_{1}-A_{2})$. Thus, for instance, if $A_{1}>A_{2}$ and $A_{1}p_{1}<A_{2}p_{2}$ this state displays backflow.

\begin{figure}[htbp]
  \begin{minipage}[t]{0.45\linewidth}
    \centering
    \includegraphics[width=\linewidth]{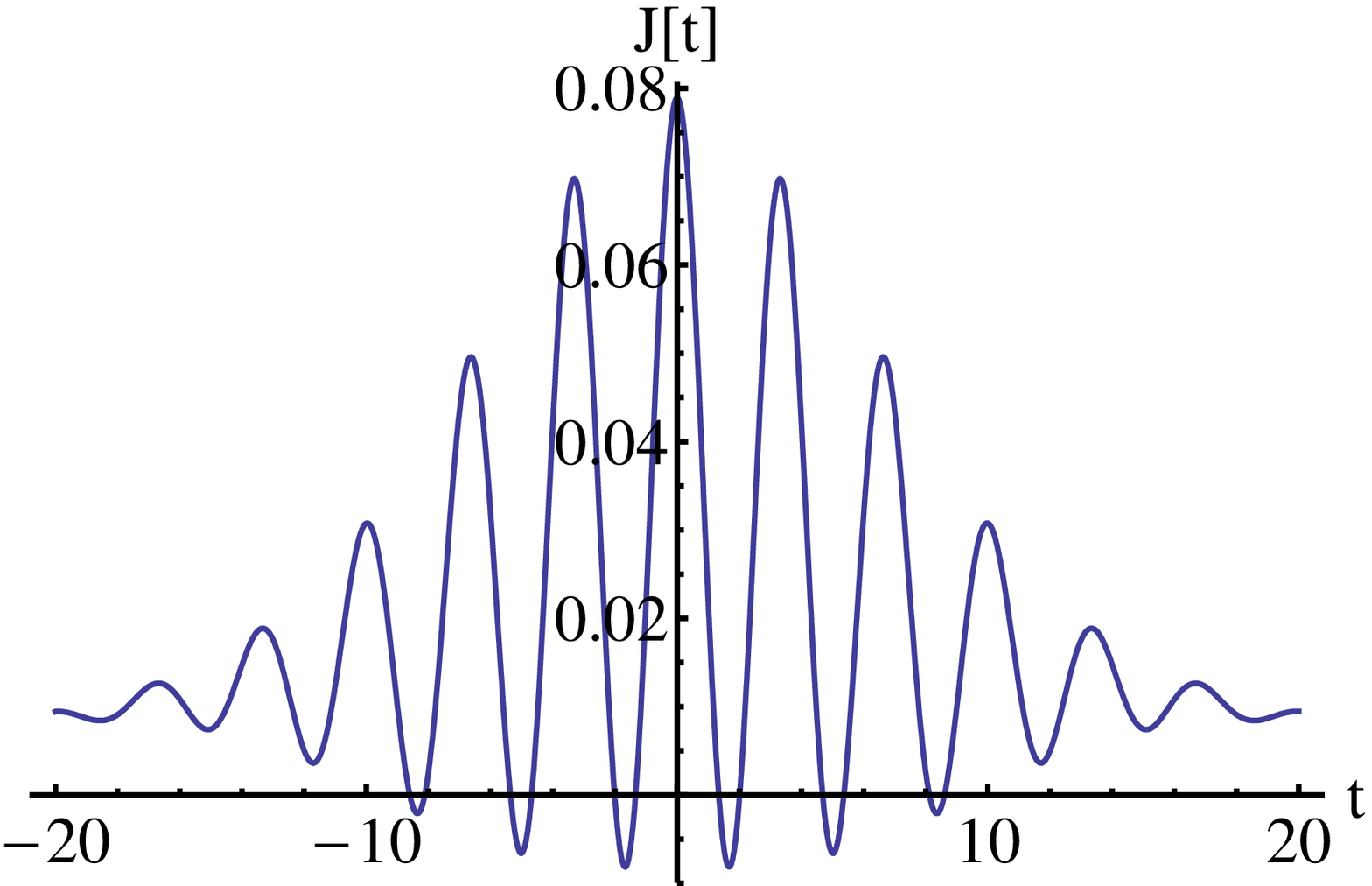}
    \caption{Plot of the current for a wavefunction consisting of a superposition of two gaussians, with the parameters given in Eq.(\ref{jsec4}).}
    \label{jfig1}
  \end{minipage}
  \hspace{1cm}
  \begin{minipage}[t]{0.45\linewidth}
    \centering
    \includegraphics[width=\linewidth]{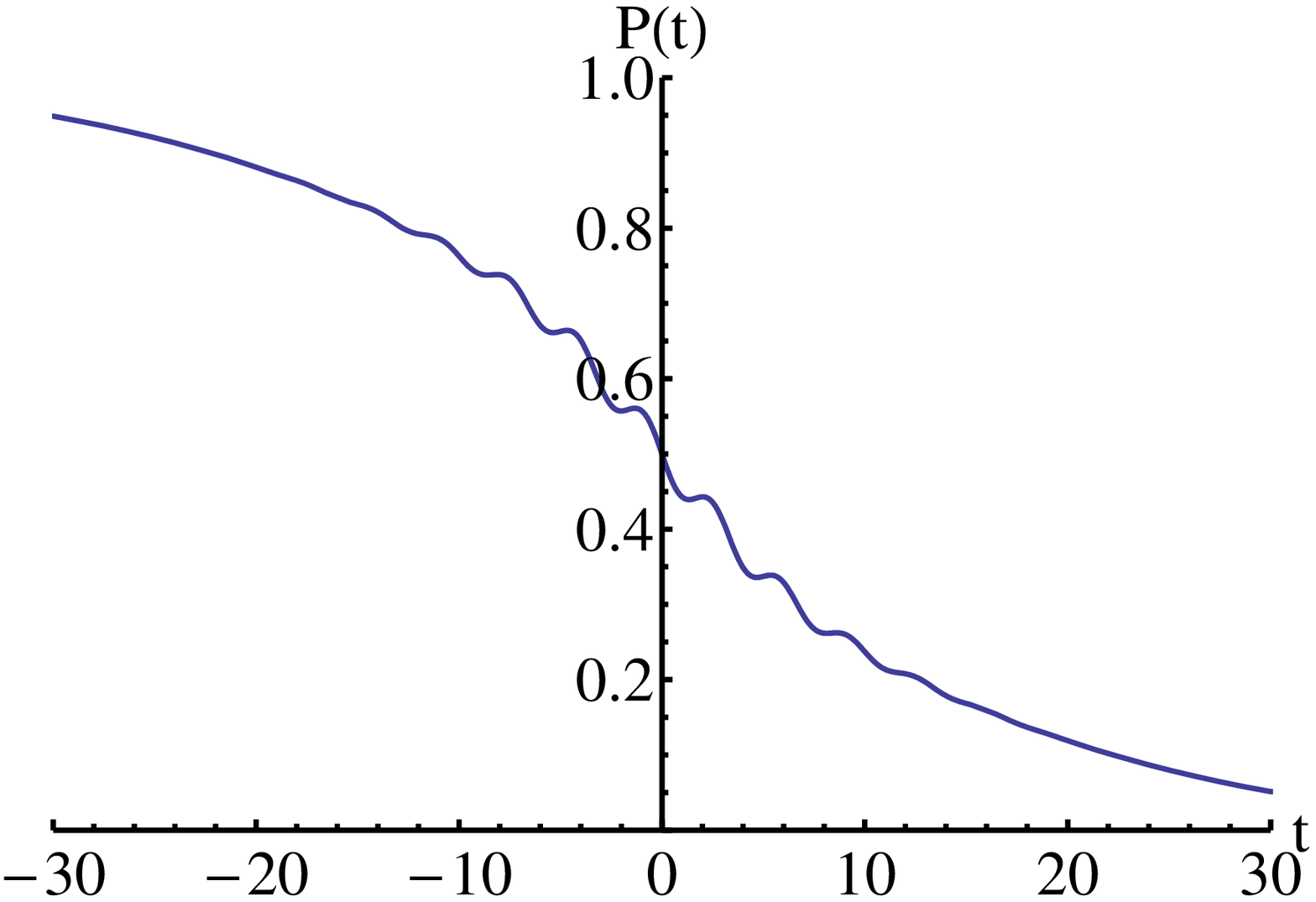}
    \caption{Plot of the probability for remaining in $x<0$ for a wavefunction consisting of a superposition of two gaussians, with the parameters given in Eq.(\ref{jsec4}).}
    \label{jfig1.1}
  \end{minipage}
\end{figure}

\begin{figure}[htbp]
  \begin{minipage}[t]{0.45\linewidth}
    \centering
    \includegraphics[width=\linewidth]{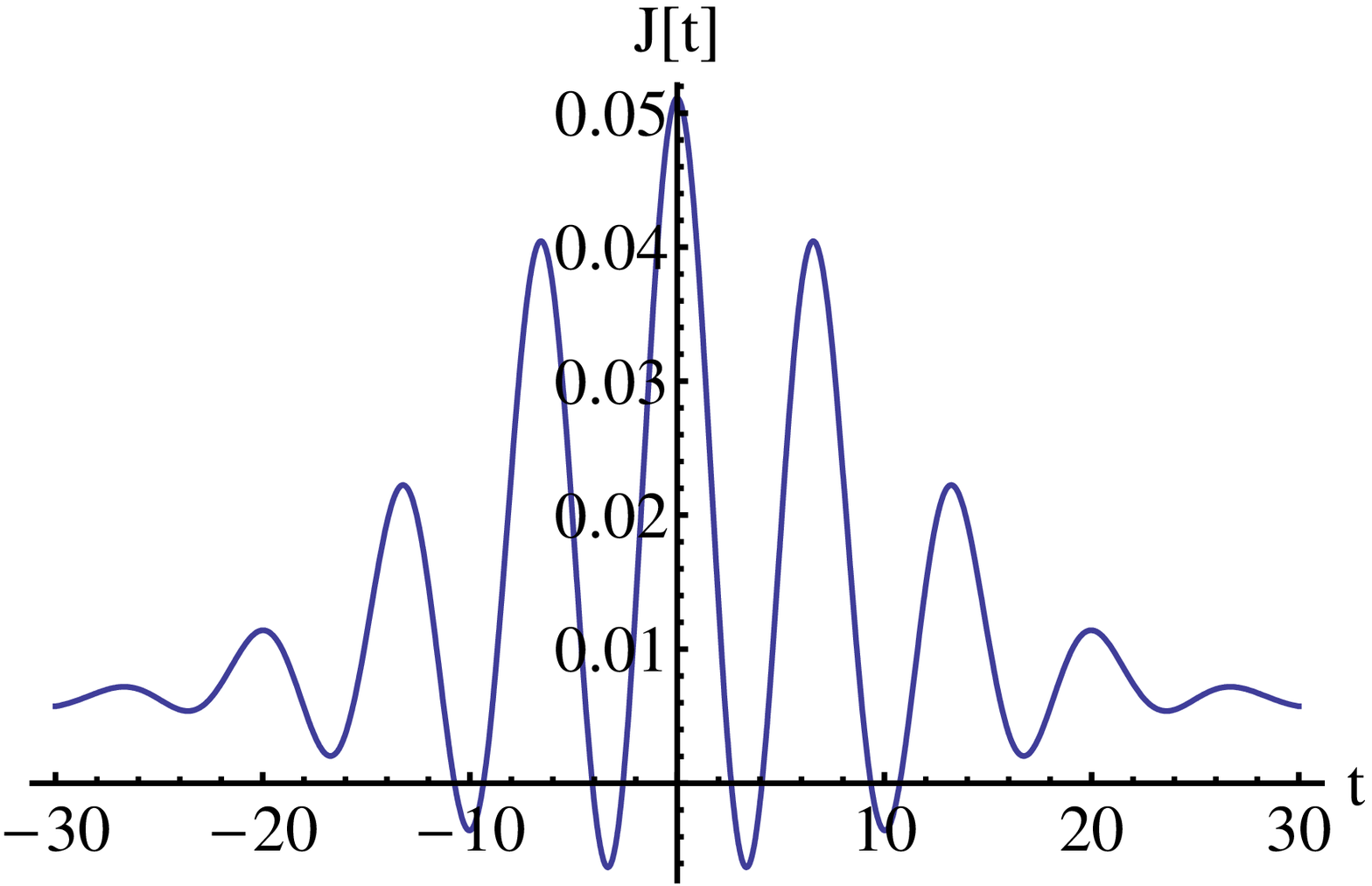}
    \caption{Plot of the current for a wavefunction consisting of a superposition of two gaussians, with the parameters given in Eq.(\ref{jsec5}).}
    \label{jfig2}
  \end{minipage}
  \hspace{1cm}
  \begin{minipage}[t]{0.45\linewidth}
    \centering
    \includegraphics[width=\linewidth]{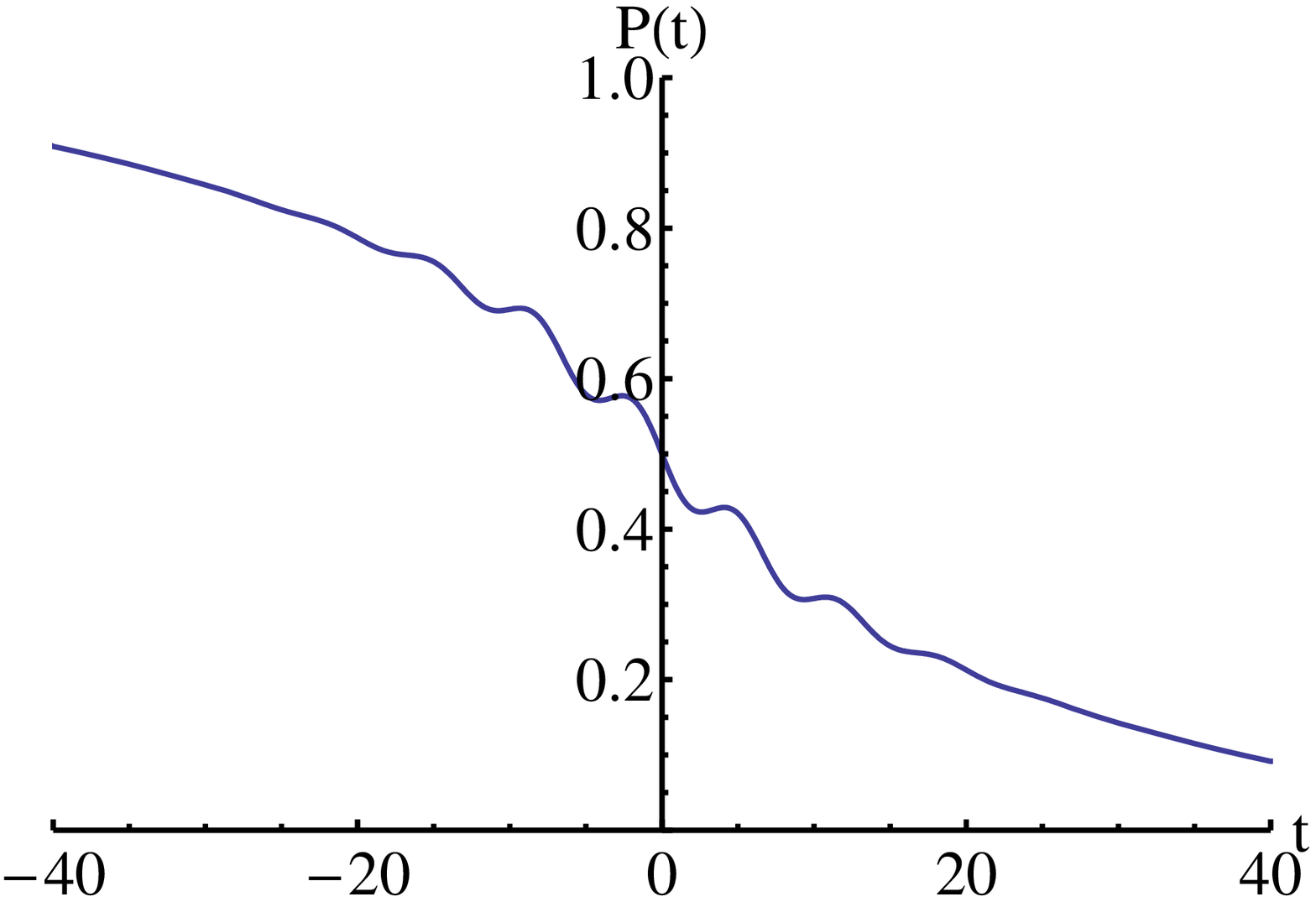}
    \caption{Plot of the probability for remaining in $x<0$ for a wavefunction consisting of a superposition of two gaussians, with the parameters given in Eq.(\ref{jsec5}).}
    \label{jfig2.1}
  \end{minipage}
\end{figure}

Consider now the normalised state
\beq
\psi(x,t)=\sum_{k=1,2}A_{k}\frac{1}{\sqrt{4\s^{2}+2it}}\exp\left(i p_{k}(x-p_{k}t)-\frac{(x-p_{k}t)^{2}}{4\s^{2}+2it}\right)\label{jsec3}.
\eeq
This is a sum of two initial gaussian wavepackets with equal spatial width $\s$, evolved for a time $t$. If we let $\s\to\infty$ we essentially recover Eq.(\ref{jsec1}). The idea is that if we take $\s$ to be large enough, the current at the origin is the product of Eq.(\ref{jsec2}) and a slowly varying function, so that the conclusions about backflow still hold. The analytic expression for the current is somewhat long and complex, and we will not give it here. Instead we show below two sets of plots of the current at the origin and the probability of remaining in $x<0$ as functions of time for the state in Eq.(\ref{jsec3}) for the following two sets of parameters, obtained by a search of parameter space of examples illustrating the backflow effect as well as possible.
In Figs.(\ref{jfig1}) and (\ref{jfig1.1}) we plot the parameters
\beq
p_{1}= 0.5, \quad p_{2}=2,\quad \s=10,\quad A_{1}=1.7,\quad A_{2}=1.\label{jsec4}
\eeq
and in Figs.(\ref{jfig2}) and (\ref{jfig2.1}) we plot the parameters
\beq
p_{1}=0.3, \quad p_{2}=1.4,\quad \s=10,\quad A_{1}=1.8,\quad A_{2}=1.
\label{jsec5}
\eeq

\begin{figure}[htbp]
  \begin{minipage}[t]{0.45\linewidth}
    \centering
    \includegraphics[width=\linewidth]{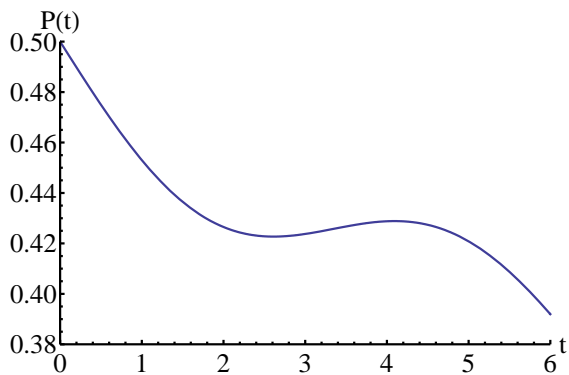}
    \caption{Close up of Fig.\ref{jfig2.1}. The probability is clearly seen to increase between $t\approx2$ and $t\approx4$.}
    \label{jfig2.2}
  \end{minipage}
  \hspace{1cm}
  \begin{minipage}[t]{0.45\linewidth}
  \end{minipage}
\end{figure}

We clearly see from these plots that there are several intervals during which the current is negative. These examples show that the backflow can occur in several disjoint time intervals.
A magnification of one of these backflow region is shown in Fig.(\ref{jfig2.2}).
The effect is robust with respect to small changes of the parameters, Eq.(\ref{jsec4}) or Eq.(\ref{jsec5}), which were chosen because they give reasonable amounts of backflow.


The set of parameters in Eq.(\ref{jsec5}) give rise to the greatest amount of backflow we have been able to find, although we have not performed anything like a comprehensive search of the parameter space.  The value of the flux during the largest period of backflow is
\beq
F=\int_{t_{1}}^{t_{2}}dt J(t),
\eeq
where $J(t)$ is the current, Eq.(\ref{cur}), and the interval $[t_{1},t_{2}]$ is chosen such that the current is negative for the whole of this time. Computing the flux during the largest backflow interval gives,
\beq
F\approx -0.0061,\label{jsec6}
\eeq
or about 16 percent of the theoretical maximum.

It is important to check that this probability backflow cannot be explained by the tiny probability of having negative momentum which comes from this gaussian state. An order of magnitude estimate will suffice here. We have two gaussian wavepackets centered about different momenta. Consider the wavepacket centered around $p=0.3$. The probability that a measurement of the momentum of this state would yield a negative answer is given approximately by,
\beq
\mbox{Prob}(p<0)\sim\int_{-\infty}^{0}dp \exp(-200(p-0.3)^{2})\sim 10^{-10}.
\eeq
The negative flow of probability is therefore entirely due to the backflow effect.

\section{An approximation to the backflow maximising state}

Backflow states may be found by solving the eigenvalue equation Eq.(\ref{evalue2}).
The numerical work of Penz et al \cite{Penz} yielded a plot of the approximate eigenstate satisfying Eq.(\ref{evalue2}) giving the most negative eigenvalue $-c_{bm}$, i.e., the largest amount of backflow. This state appears to be of the form $ \phi (u) \sim \sin u^2 / u $ for large $u$.

A problem of great interest is to find analytic expressions for wave functions with backflow which match these numerical results and asymptotic results as closely as possible and for which the current at arbitrary time may be computed analytically. This is what we do in this section. To be clear, we will not find approximate analytic solutions to the eigenvalue problem, Eq.(\ref{evalue2}). Rather, inspired by the numerical solutions to the eigenvalue problem, we will exhibit analytic expressions for wave functions, compute their current at time $t$ analytically, and show that they have significant negative flux, calculated using Eq.(\ref{flux3}).

\subsection{Numerical results}

We first review the numerical results of the computation of the backflow state.
We have repeated the numerical analysis of Penz et al.\cite{Penz} of the optimizing state and its current. This is purely for comparison with our analytic results and we do not claim any improvements in accuracy over Penz et al.

\begin{figure}[h]
\begin{center}
\includegraphics[width=5in]{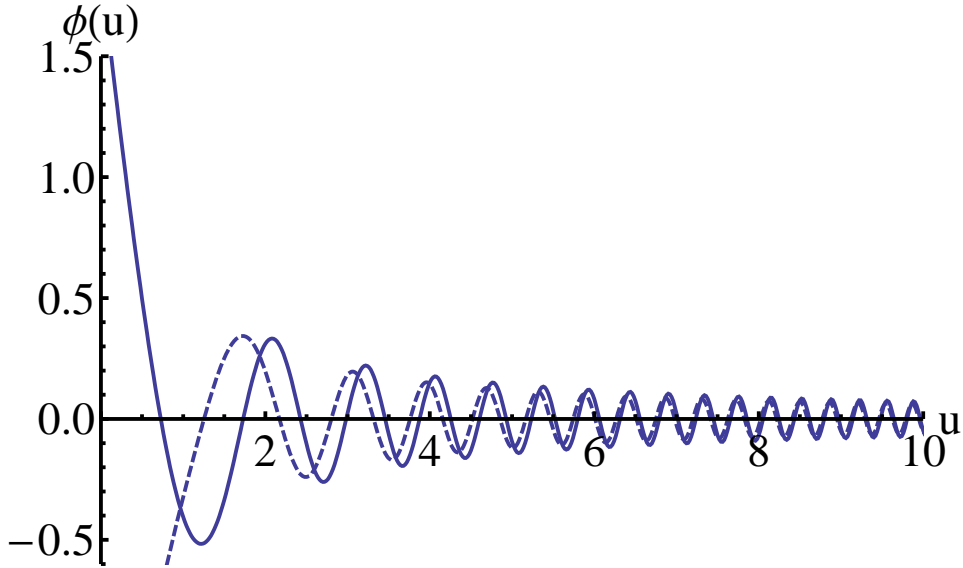}
\caption{Plot of the backflow maximising state (solid line) together with $\phi_{as}$ for $a=0$, $b=-0.1$ (dashed line).}
\label{figphiexact}
\end{center}
\end{figure}

\begin{figure}[h]
\begin{center}
\includegraphics[width=5in]{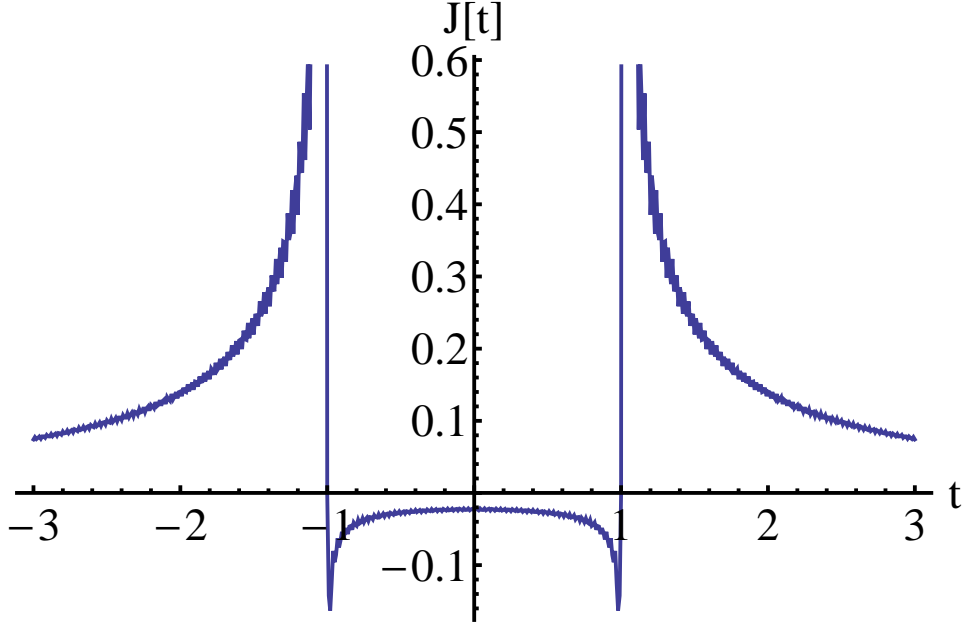}
\caption{The current, $J(t)$, as computed from the numerically obtained backflow maximizing state.}
\label{figjexact}
\end{center}
\end{figure}

Our numerically computed maximum backflow state has the asymptotic form,
\beq
\phi_{as}(u)=a \frac{\sin(u^2)}{u}+b \frac{\cos(u^{2})}{u}.\label{jsecas}
\eeq
In Fig.(\ref{figphiexact}) we plot the numerically computed maximum backflow state together with $\phi_{as}$ for $a=0$, $b=-0.1$. Comparing by eye it seems like these parameters produce the best fit, but we will consider states with the more general form Eq.(\ref{jsecas}).
We plot in Fig.(\ref{figjexact}) the current computed from the numerically obtained backflow maximizing state.
Note that the current appears to have a very specific singularity structure at $t= \pm 1$, at which it jumps between $- \infty$ and $ + \infty$. (This is presumably related to some properties of the flux operator).
These two plots are in agreement, in general shape, with the numerical results of Penz et al \cite{Penz} and we will compare our analytic results with these plots in what follows.

\subsection{The Form of the Extremizing State}

We first make a brief remark about the asymptotic form $\phi (u) \sim \sin u^2 / u $, used by
Penz et al \cite{Penz} mainly to check for square-integrability, but it is useful to check its current to see if it has backflow. It is easy to see that it has positive current at $t=0$,
contrary to the numerical result shown in Fig.(\ref{figjexact}). We see this as follows.
In terms of the dimensionless quantities introduced in Section 2, the current at $t=0$ is
\bea
J(0)&=&\frac{1}{2\pi}\int_{0}^{\infty}du dv \ (u+v)\phi(u)\phi(v)\nonumber\\
&=&\frac{1}{\pi}\int_{0}^{\infty}du\ \frac{\sin(u^{2})}{u}\int_{0}^{\infty}dv\sin(v^{2})\nonumber\\
&=&\frac{1}{8}\sqrt{\frac{\pi}{2}}>0.
\label{cur8}
\eea
Hence, to have negative current, the optimizing state must be quite different from $\sin u^2 / u $ for small $u$, but may agree with it asymptotically.

Another obvious guess for the approximate form of the extremizing state is the Bessel function $J_0 (u^2 )$. However, using the formulae
\bea
\int_0^\infty du \ u J_0 (u^2) &=& \half
\\
\int_0^\infty du \ J_0 (u^2) &=& \sqrt{2} \frac{  \Gamma (5/4) } { \Gamma (3/4) } \approx 1.04605
\eea
it is clear that the current at $t=0$, Eq.(\ref{cur8}), is again positive.

These unsuccessful guesses, combined with a process of trial and error, have led us to the two guesses for which the current at arbitrary times can be computed analytically and which have substantial backflow.
It has not been difficult to simply guess momentum space wave functions matching the numerical solution
in Fig.(\ref{figphiexact}). (However, we have found that it is difficult to match the precise singularity structure of the current at $t= \pm 1$, as we shall see).

\subsection{Guess 1}

\begin{figure}[h]
\begin{center}
\includegraphics[width=5in]{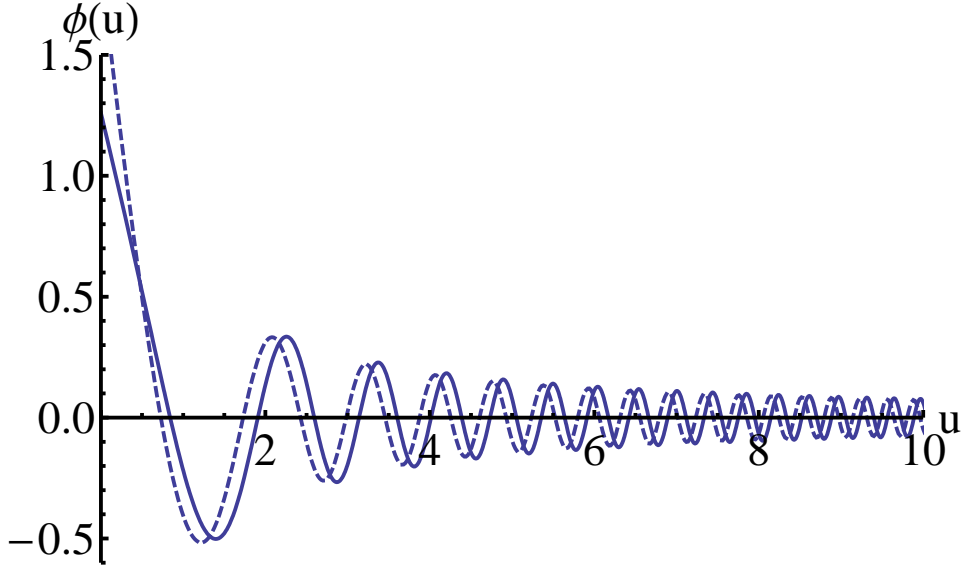}
\caption{$\phi(u)$, for a=0.4 (solid line), with the exact numerical result for comparison (dashed line).}
\label{phi3}
\end{center}
\end{figure}

Our first guess is the momentum space wavefunction $\phi_{1}(u)$ given by,
\beq
\phi_{1}(u)=N\left[(\frac{1}{2}-C(u))+a(\frac{1}{2}-S(u))\right],\quad a\in \mathbb{R}
\eeq
where $N$ is some normalisation factor. Here
\beq
C(u)=\mbox{FresnelC}\left(\sqrt{{\frac{2}{\pi}}}u\right)=\sqrt{\frac{2}{\pi}}\int_{0}^{u}dx\cos(x^{2})
\eeq
and similarly for $S(u)$.
This state has the asymptotic form,
\beq
\phi_{1}(u)\sim N\left[-\frac{\sin(u^{2})}{u}+a \frac{\cos(u^{2})}{u}\right].
\eeq
Note that,
\beq
\phi_{1}(u)=N\sqrt{\frac{2}{\pi}}\int_{1}^{\infty}dz u\left(\cos(z^{2} u^{2})+a\sin(z^{2}u^{2})\right)\label{1.2}
\eeq
which is a form we shall use below. Note that the norm $N$ is given by,
\beq
N^{-2}=\frac{1}{4\sqrt{\pi}}(1+a^{2}+2a(1+a)(\sqrt{2}-1))
\eeq
We plot the wavefunction in Fig.(\ref{phi3}), and we see that it shows reasonable agreement with the numerical result.

We wish to evaluate the flux, Eq.(\ref{flux3}) which may be written
\beq
F=\int_{-1}^{1}dt \ J(t)\label{jsecflux0}
\eeq
where we have introduced the current
\beq
J(t)=\frac{1}{2\pi}\int_{0}^{\infty}du dv (u+v)\exp(i t(u^{2}-v^{2}))\phi(u)\phi(v)\label{jsecj}
\eeq
written in terms of the dimensionless variables introduced in Section 2 (and here $t$ is a dimensionless time parameter).

Because we are working only with approximate eigenstate of the flux, the interval during which the $J(t)$ is negative may not coincide exactly with $[-1,1]$, so for that reason, we take the flux instead to be
\beq
F=\int_{t_1}^{t_2}dt \ J(t)\ \label{jsecflux}
\eeq
and adjust the interval $[t_1,t_2]$ to match the region of negative flux.
Eq.(\ref{jsecj}) can also be written as,
\bea
J(t)&=&\mbox{Re}\left(\frac{1}{\pi}\int_{0}^{\infty}du\exp(i tu^{2})\phi(u)\int_{0}^{\infty}dv\; v \exp(-itv^{2})\phi(v)\right)\nonumber\\
&=&\frac{1}{\pi}\mbox{Re}\left(U(t) V(t)\right)\label{jsecu},
\eea
so that we can compute the current by first computing each of the integrals in Eq.(\ref{jsecu}) seperately.

We begin by computing the $U$ integral in Eq.(\ref{jsecu}). We use Eq.(\ref{1.2}) to write
\beq
U_{1}(t)=N\sqrt{\frac{2}{\pi}}\int_{0}^{\infty}du\int_{1}^{\infty}dz  \frac{u}{2}\left((1-ia)\exp(i(t+z^{2})u^{2})+(1+ia)\exp(i(t-z^{2})u^{2})\right)
\eeq
We would like to change the order of integration at this point, but we cannot, since the $u$ integral only converges conditionally. To remedy this we introduce a convergence factor $\exp(-\e u^{2})$ where $\e>0$. We can then write,
\bea
U_{1}(t)&=&N\sqrt{\frac{2}{\pi}}\int_{1}^{\infty}dz\int_{0}^{\infty}du  \frac{u}{2}\nonumber\\
&&\times\Big[(1-ia)\exp(i(t+z^{2})u^{2}-\e u^{2})+(1+ia)\exp(i(t-z^{2})u^{2}-\e u^{2})\Big]\nonumber\\
&=&\frac{N}{2\sqrt{2\pi}}\int_{1}^{\infty}dz\left(\frac{1-ia}{\e-i(t+z^{2})}+\frac{1+ia}{\e-i(t-z^{2})}\right)\nonumber\\
&=&\frac{iN}{2\sqrt{2\pi}}\left[(1-ia)\frac{\mbox{ArcTan}(\frac{z}{\sqrt{t+i\e}})}{\sqrt{t+i\e}}- (1+ia)\frac{\mbox{ArcTan}(\frac{z}{\sqrt{-t-i\e}})}{\sqrt{-t-i\e}}\right]_{z=1}^{\infty}\nonumber\\
&=&\frac{N}{2\sqrt{2\pi(t+i\e)}}\left[\frac{\pi(1+i)(1+a)}{2}-(i+a)\mbox{ArcTan}(\frac{1}{\sqrt{t+i\e}})\right.\nonumber\\
&&\left.-(i-a)\mbox{ArcTanh}(\frac{1}{\sqrt{t+i\e}})\right]
\eea
where we have used the standard integrals,
\bea
\int du \frac{1}{\a+u^{2}}&=&\frac{\mbox{ArcTan}(\frac{u}{\sqrt{\a}})}{\sqrt{\a}}\\
\int du \frac{1}{\a-u^{2}}&=&\frac{\mbox{ArcTanh}(\frac{u}{\sqrt{\a}})}{\sqrt{\a}}
\eea

We now turn to the $V$ integral in Eq.(\ref{jsecu}). Again, we use Eq.(\ref{1.2}) to write,
\bea
V_{1}(t)&=&N\sqrt{\frac{2}{\pi}}\int_{0}^{\infty}dv \int_{1}^{\infty}dz  \frac{v^{2}}{2}\nonumber\\
&&\times\Big[((1-ia)\exp(-i(t-z^{2})v^{2})+(1+ia)\exp(-i(t+z^{2})v^{2})\Big]
\eea
As before, in order to change the order of integration we insert a convergence factor,
\bea
V(t)_{1}&=&N\sqrt{\frac{2}{\pi}}\int_{1}^{\infty}dz\int_{0}^{\infty}dv  \frac{v^{2}}{2}\nonumber\\
&&\times\Big[(1-ia)\exp(-i(t-z^{2})v^{2}-\e v^{2})+(1+ia)\exp(-i(t+z^{2})v^{2}-\e v^{2})\Big]\nonumber\\
&=&\frac{N}{4\sqrt{2}}\int_{1}^{\infty}dz \left(\frac{1-ia}{(\e+i(t-z^{2}))^{3/2}}+\frac{1+ia}{(\e+i(t+z^{2}))^{3/2}}\right)\nonumber\\
&=&\frac{N}{4\sqrt{2}}\left[(1-ia)\frac{z\sqrt{-i}\sqrt{z^{2}-t+i\e}}{(t-i\e)(z^{2}-t+i\e)}-(1+ia)\frac{z\sqrt{i}\sqrt{z^{2}+t-i\e}}{(t-i\e)(z^{2}+t-i\e)}\right]_{1}^{\infty}\nonumber\\
&=&\frac{N}{4\sqrt{2}}\left[(1-ia)\frac{\sqrt{-i}}{t-i\e}\left(1-\frac{1}{\sqrt{1-t+i\e}}\right)\right.\nonumber\\
&&\left.-(1+ia)\frac{\sqrt{i}}{t-i\e}\left(1-\frac{1}{\sqrt{1+t-i\e}}\right)\right]\nonumber\\
\eea
where we have used the standard integrals,
\bea
\int du \frac{1}{(a+u^{2})^{3/2}}&=&\frac{u}{a\sqrt{u^{2}+a}}\\
\int du \frac{1}{(a-u^{2})^{3/2}}&=&\frac{u}{a\sqrt{u^{2}-a}}
\eea

\begin{figure}[h]
\begin{center}
\includegraphics[width=5in]{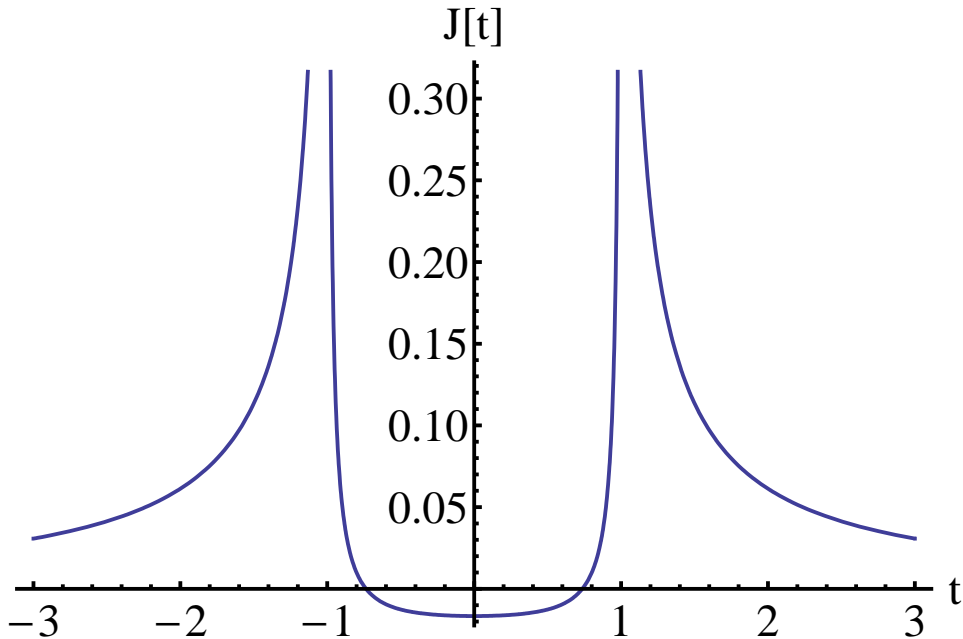}
\caption{Current, $J_{1}(t)$, for $a=0.4$ and $\e=10^{-7}$.}
\label{j3}
\end{center}
\end{figure}

Given $U$ and $V$ we may now plot the current, Eq.(\ref{jsecu}). The plot is shown in Fig.(\ref{j3}). It has a substantial period of backflow and is in broad agreement with
the numerical result Fig.(\ref{figjexact}), although differs significantly in the behavior near $t=\pm1$.  We have chosen the value $a=0.4$ which approximately maximises the backflow for this wavefunction. The flux, Eq.(\ref{jsecflux}), for this choice may be calculated by numerically integrating the current and is approximately
\beq
F=-0.02095 
\eeq
The amount of negative flux obtained is of the order of 55 percent of $c_{bm}$, which is a much greater fraction than we were able to achieve in Section 3 using superpositions of gaussians.

\subsection{Guess 2}

\begin{figure}[h]
\begin{center}
\includegraphics[width=5in]{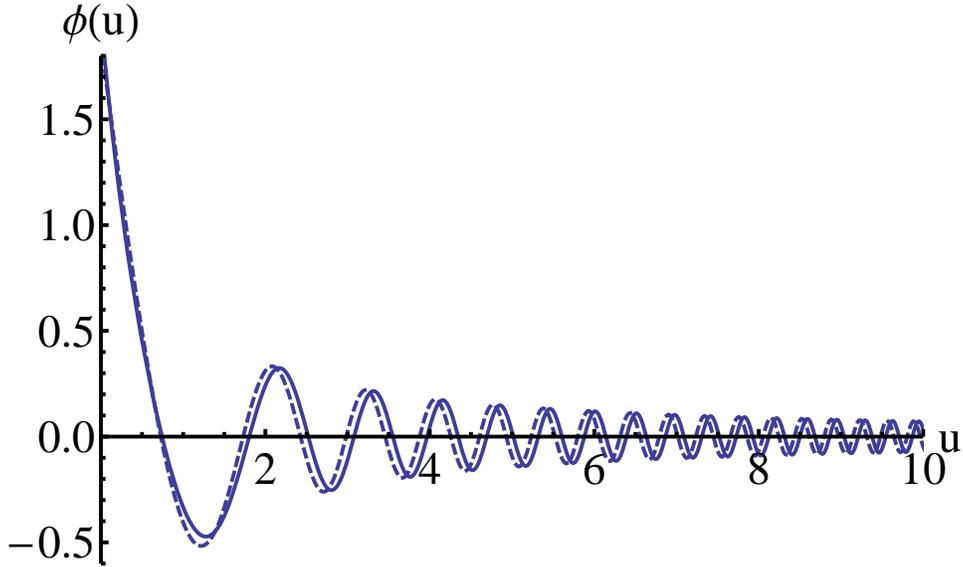}
\caption{$\phi(u)$, for a=0.6 and b=2.8 (solid line), with the numerical result for comparison (dashed line).}
\label{phi2.1}
\end{center}
\end{figure}

Our second guess is the momentum space wavefunction
\beq
\phi_{2}(u)=N\left[a e^{-b u}+(\frac{1}{2}-C(u))\right],\quad a,b\in\mathbb{R}
\label{guess2}
\eeq
where $N$ is a normalisation factor. This has the asymptotic form,
\beq
\phi_2(u)\sim N \frac{\sin(u^2)}{u}.
\eeq
We plot $\phi_{2}$ in Fig.(\ref{phi2.1}) for the values of $a$ and $b$ which maximize backflow. We see good agreement with the numerical result.

We first calculate the norm $N$,
\beq
N^{-2}=\int_{0}^{\infty}du\left(a^{2}e^{-2 b u}+(\frac{1}{2}-C(u))^{2}+2a e^{-b u}(\frac{1}{2}-C(u))\right).
\eeq
The first two terms can be computed easily enough. The last term is more challenging. Using the standard integral,
\beq
\int_{0}^{\infty}dx e^{-ax }C(x)=\frac{1}{a}\left\{\left[\frac{1}{2}-S\left(\frac{a}{\pi}\right)\right]\cos\left(\frac{a^{2}}{2}\right)-\left[\frac{1}{2}-C\left(\frac{a}{\pi}\right)\right]\sin\left(\frac{a^{2}}{2}\right)\right\}
\eeq
we find,
\bea
N^{-2}&=&\frac{a^{2}}{2b}+\frac{1}{4\sqrt{\pi}}+\frac{a}{b}-\nonumber\\
&&\frac{2a}{b}\left\{\left[\frac{1}{2}-S\left(\frac{b}{\pi}\right)\right]\cos\left(\frac{b^{2}}{2}\right)-\left[\frac{1}{2}-C\left(\frac{b}{\pi}\right)\right]\sin\left(\frac{b^{2}}{2}\right)\right\}
\eea


We now turn to computing the $U$ integral in Eq.(\ref{jsecu}). We wish to compute
\bea
U_{2}(t)&=&N\int_{0}^{\infty}du e^{it u^{2}-\e u^{2}}\left(a e^{-b u}+\frac{1}{2}-C(u)\right)\nonumber\\
&=&aN\int_{0}^{\infty} due^{-bu-(\e-it)u^{2}}+N\int_{0}^{\infty}du e^{-(\e-it)u^{2}}(\frac{1}{2}-C(u))
\eea
where we have added a convergence factor $e^{-\e u^{2}}$. The second integral we have already seen, it is just $U_{1}(t)$ for out first guess of $\phi_{1}(u)$, with the coefficient of the term involving $S(u)$ taken to be zero. The first integral can also be done easily, using the standard integral
\beq
\int_{0}^{\infty}dx \exp(-\a x^{2}-\b x)=\frac{1}{2}\sqrt{\frac{\pi}{\a}}e^{\frac{\b^{2}}{4\a}}\mbox{Erfc}\left(\frac{\b}{2\sqrt{\a}}\right), \quad \mbox{for Re}(\a)>0.
\eeq
Combining these gives,
\bea
U_{2}(t)&=&\frac{aN}{2}\sqrt{\frac{i\pi}{t+i\e}}e^{\frac{ib^{2}}{4(t+i\e)}}\mbox{Erfc}\left(\frac{b}{2}\sqrt{\frac{i}{t+i\e}}\right)\nonumber\\
&&+\frac{N}{2\sqrt{2\pi(t+i\e)}}\left[\frac{\pi(1+i)}{2}-i\mbox{ArcTan}(\frac{1}{\sqrt{t+i\e}})-i\mbox{ArcTanh}(\frac{1}{\sqrt{t+i\e}})\right]
\eea


Next we compute the $V$ integral in Eq.(\ref{jsecu}),
\bea
V_{2}(t)&=&N\int_{0}^{\infty}dv v e^{-(\e+it) v^{2}}\left(a e^{-b v}+\frac{1}{2}-C(v)\right)\nonumber\\
&=&aN\int_{0}^{\infty}dv v e^{-b v-(\e+
it)v^{2}}+\int_{0}^{\infty}dv ve^{-(\e+it)v^{2}}(\frac{1}{2}-C(v))
\eea
As with $U_{2}(t)$, the second integral can be simply written down by comparing to $V_{1}(t)$ for our first guess of $\phi_{1}(u)$. The first integral is also easily performed, using the standard integral,
\beq
\int_{0}^{\infty}dx x e^{-\a x^{2}-\b x}=\frac{1}{2\a}\left(1-\frac{\b}{2}\sqrt{\frac{\pi}{\a}}\exp\left(
-\frac{\b^{2}}{4\a}\right)\mbox{Erfc}\left(\frac{\b}{2\sqrt{a}}\right)\right), \quad \mbox{for Re}(\a)>0.
\eeq
We thus find,
\bea
V_{2}(t)&=&\frac{aN}{2(\e+it)}\left(1-\frac{b}{2}\sqrt{\frac{\pi}{\e+it}}\exp\left(
-\frac{b^{2}}{4(\e+it)}\right)\mbox{Erfc}\left(\frac{b}{2\sqrt{\e+it}}\right)\right)\nonumber\\
&&+\frac{N}{4\sqrt{2}}\left[\frac{\sqrt{-i}}{t-i\e}\left(1-\frac{1}{\sqrt{1-t+i\e}}\right)-\frac{\sqrt{i}}{t-i\e}\left(1-\frac{1}{\sqrt{1+t-i\e}}\right)\right]
\eea

\begin{figure}[ht]
\begin{center}
\includegraphics[width=5in]{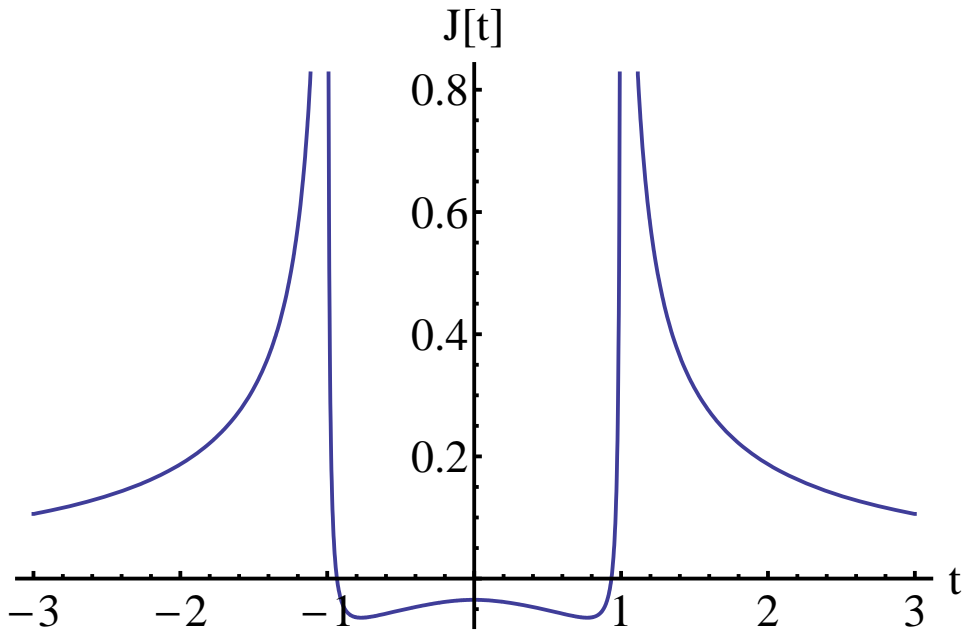}
\caption{Current, $J_{2}(t)$ for $a=0.6$, $b=2.8$, and $\e=10^{-7}$.}
\label{J2}
\end{center}
\end{figure}

Now we have $U$ and $V$ we can compute the current Eq.(\ref{jsecu}) and, by numerical integration, the flux Eq.(\ref{jsecflux}). The maximum amount of negative flux we can generate is
\beq
F=-0.02757,
\eeq
which occurs for the parameters $a=0.6$, $b=2.8$. This corresponds to about 70 percent of the maximum
$c_{bm}$. We plot the current $J(t)$ for these parameters in Fig.(\ref{J2}). The current is closer to the numerical result, Fig.(\ref{figjexact}) than our first guess, Fig.(\ref{j3}) but still lacks the correct behavior as $t\to\pm1$.

\begin{figure}[ht]
\begin{center}
\includegraphics[width=5in]{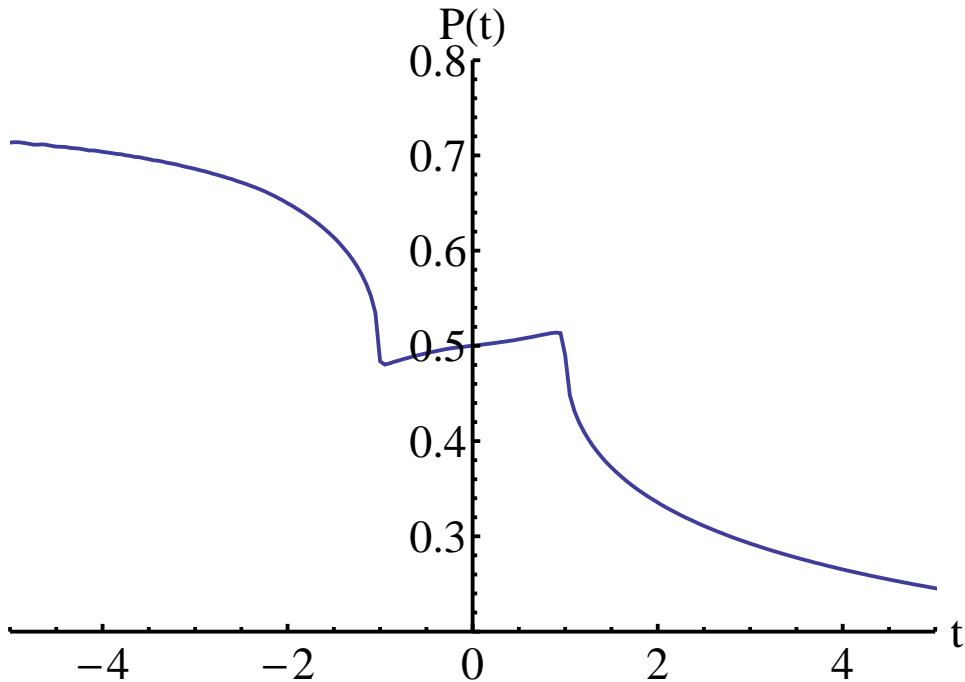}
\caption{Probability $P(t)$  that the state will be found in $x<0$ at time $t$ for the wavefunction $\phi_{2}$. The probability is clearly seen to increase during the interval $[-1,1]$.}
\label{figptapprox}
\end{center}
\end{figure}

Finally, to give a clear illustration of the backflow phenomenon, in Fig.(\ref{figptapprox}) we plot the probability that the state will be found in $x<0$ as a function of time, for the state Eq.(\ref{guess2}). The probability decreases over the whole interval plotted, but has a very noticeable period of increase between $t=\pm1$.


\section{The Classical Limit of Backflow}

Some insights into the properties of backflow may be found by looking at its classical
limit. In the usual account of emergent classicality for the free particle considered here, one considers a larger system in which the particle is coupled to a wider environment and one considers the evolution of the reduced density matrix of the particle only \cite{JoZe,Hal8}. It is well-known that, in a wide variety of such open system models, the Wigner function of the particle will become positive after a short period of time \cite{DiKi}. From Eq.(\ref{Wig}), it is then easily seen that the flux will then be positive, so the backflow clearly goes away in the standard approach to the classical limit. A detailed discussion of the arrival time problem in the presence of an environment was given in Ref.\cite{Ye1} and in this work the resulting positivity of the current, after finite time, is clearly seen.

However, there is a more interesting and subtle question, noted by Bracken and Melloy \cite{BrMe},
which is that the eigenvalues of the flux operator are independent of $\hbar $, as we saw in Eq.(\ref{evalue2}), despite the fact that the existence of negative eigenvalues (negative flux) is clearly a quantum phenomenon. This means that in the naive classical limit,
$\hbar \rightarrow 0$, the backflow does not go away, as one might expect. Of course, this ``limit'' is an oversimplification of what the classical limit means, but despite this, it is still disconcerting that this obviously quantum phenomenon is apparently independent of $\hbar$.

This situation is reminiscent of another situation without naive classical limit, namely, scattering off a step potential,
where it is known that the quantum-mechanical reflection coefficient is independent of $\hbar$, so does not go to zero as $\hbar \rightarrow
0$, contrary to classical expectations. The origin of the problem is the use of an
exact step potential. If instead a smoothed off step is used, with lengthscale $\sigma$
describing the size of the smoothing region, then reflection does indeed go away
if $\hbar \rightarrow 0 $ with $\sigma$ held constant and non-zero \cite{LaLi}.
The point here is that the exact step potential is an idealization that fails to capture all physical properties. Replacement with a more realistic potential restores the naive classical limit.

In the backflow situation, we may therefore also expect to get a reasonable naive classical
limit by small modification of the situation. In particular, instead of defining
the flux operator in terms of exact projection operators $P = \theta (\hat x)$, we define it in terms of a quasiprojector $Q$. This seems reasonable since, as discussed earlier, backflow can be measured by measuring whether the particle is in $x>0$ at two different times and, due to the inevitable imprecision of real measurements, such measurements are best modeled by quasiprojectors.
A convenient choice of quasiprojector is
\beq
Q = \int_0^{\infty} dy \ \delta_{\sigma} ( \hat x - y )
\label{quasi}
\eeq
where $\delta_{\sigma} (\hat x - y ) $ is a smoothed out $\delta$-function,
\beq
\delta_{\sigma} (\hat x - y ) = \frac {1} { (2 \pi \sigma^2)^{1/2} }
\exp \left( - \frac { (\hat x - y)^2 } {2 \sigma^2} \right)
\eeq
This goes to the usual $\delta$-function as $\sigma \rightarrow 0 $ and then $Q \rightarrow \theta (\hat x ) $. If we replace $P$ with the quasiprojector $Q$ in the expression for the flux, we get
Eq.(\ref{flux2}) but with the current operator replaced by
\beq
\hat J = \frac {1} {2 m} \left( \hat p \delta_{\sigma} (\hat x) + \delta_{\sigma} (\hat x) \hat p \right)
\label{curop2}
\eeq
The resulting flux will, loosely speaking, by less negative, since the commutator between $\hat p$ and $\delta_\sigma (\hat x )$ becomes smaller as $\sigma $ becomes larger.

With the quasiprojector, the flux written in the form Eq.(\ref{flux3}) acquires an exponential factor
\beq
F(-T/2,T/2) =  \frac {1} {\pi}\int_0^\infty du  \int_0^\infty dv \ \phi^*(u) \  \frac { \sin (u^2 - v^2)  } {(u-v)} e^{ -a^2 (u-v)^2} \ \phi (v)
\label{flux4}
\eeq
where the dimensionless number $a$ is given by $ a^2 = 2 m \sigma^2 / \hbar T$ and the eigenvalue equation Eq.(\ref{evalue2}) will acquire the same exponential factor,
\beq
\frac {1} {\pi} \int_0^\infty dv \ \frac { \sin (u^2 - v^2)  } {(u-v)}\ e^{ -a^2 (u-v)^2} \
\phi (v)
 = \lambda \phi (u)
\label{evalue4}
\eeq
This means that the eigenvalues $\lambda$ will now depend on $a$, so we write $\lambda = \lambda (a)$. Through $a$ they will therefore depend on $\hbar$ and the ``limit'' $\hbar \rightarrow 0 $ now clearly means the regime $ a \gg 1 $, that is, $ \hbar \ll 2 m \sigma^2 / T $.
Hence, in a more realistic measurement situation, the bound on the total backflow -- the most negative eigenvalue of Eq.(\ref{evalue4}) -- {\it will} depend on $\hbar$ and the limit $\hbar \rightarrow 0 $ may now be more meaningful.

Bracken and Melloy noticed a similar phenomenon in two other models. Firstly, in the context of the Dirac equation, where the presence of the speed of light as another physical parameter permits the construction of a dimensionless parameter analogous to $a$ above \cite{BrMe2}. Secondly, in a non-relativistic model with a constant force, which again introduces a new physical parameter \cite{BrMe3}.

A reasonable conjecture is that the negative eigenvalues will increase with $a$ and also that
\beq
\lambda (a) \ge - c_{bm}
\eeq
for all $a$, so that the Bracken-Melloy bound $-c_{bm}$ emerges as a lower bound on the eigenvalues, achievable only in the limit $a \rightarrow 0 $. It seems unlikely, however, that all the negative eigenvalues will all become positive or zero, except perhaps in the limit $a \rightarrow \infty $. This behaviour is best explored numerically, which we now consider.

\begin{figure}[ht]
\begin{center}
\includegraphics[width=5in]{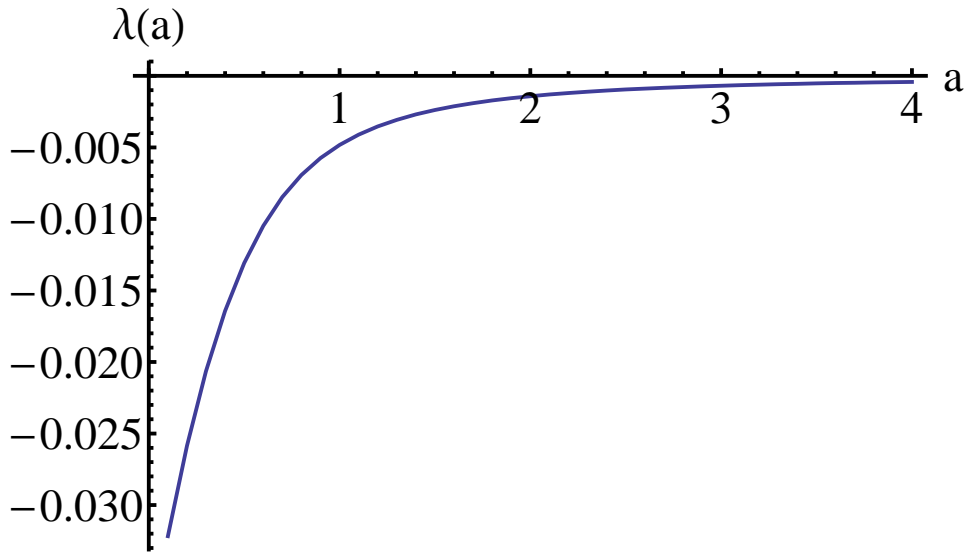}
\caption{Plot of the size of the most negative eigenvalue of Eq.(\ref{evalue4}), $\lambda(a)$ as a function of $a$. }
\label{lambdaplot}
\end{center}
\end{figure}

We consider the behaviour of the most negative eigenvalue $\lambda (a)$ of Eq.(\ref{evalue4}). We have not been able to solve this equation analytically, so instead we have obtained numerical estimates for $\lambda(a)$ for various values of $a$, and we plot the result in Fig.(\ref{lambdaplot}). The value of $\lambda(a)$ does indeed increase with $a$, tending to zero asymptotically. In fact, numerical solutions are consistent with the asymptotic form,
\beq
\lambda(a)\sim \ - \frac{1} {a^{2}} \label{lambdaas}
\eeq
for large $a$. One can get some analytic evidence for this result from the eigenvalue equation Eq.(\ref{evalue4}) in the limit of large $a^2$, which is
\beq
\frac {1} {\pi} \int_0^\infty dv \ (u+v)  e^{ -a^2 (u-v)^2} \
\phi (v)
 \approx \lambda (a) \phi (u)
\label{evalue5}
\eeq
since different values of $u$ and $v$ are suppressed by the exponential. By simple scaling of $u$ and $v$, it is easily seen that if $ \phi_0 (u) $ is an eigenstate of Eq.(\ref{evalue5}) with eigenvalue
$\lambda (a_0)$, then $\phi (u) = \phi_0 ( u a /a_0) $ is an eigenstate with eigenvalue $\lambda (a)
= (a_0/a)^2 \lambda (a_0)$. Hence the eigenvalues scale like $1/a^2$ for large $a$.

To confirm that this result about the negative eigenvalues is significant, we need also to compare it with the behaviour of the positive eigenvalues for large $a$. (It could be, for example, that they all go to zero, rendering the above result spurious). As noted in Section 2, wave packets which clearly cross or do not cross the origin during the interval $[-T/2,T/2]$ are approximate eigenstates of Eq.(\ref{evalue2}) with eigenvalues $1$ or $0$. It is reasonably clear that they will also be approximate eigenstates of the modified eigenvalue equation Eq.(\ref{evalue4}) as long as the wave packets cross the origin sufficiently far from the end-points of the interval $[-T/2,T/2]$, since under these conditions the incoming wave packet does not notice the smearing of the projector into a quasiprojector. This is also backed up by numerical work. We have computed a number of eigenvalues for different values of $a$. There appears to be a reasonably even distribution of positive eigenvalues in the interval $[0,1]$ for a range of values of $a$, so, unlike the negative eigenvalues, this part of the spectrum is not significantly changed as $a $ becomes large.

In summary, for smeared projectors, Eq.(\ref{quasi}), which are perhaps better models for real measurements than exact projectors, the eigenvalues and in particular the lowest eigenvalue representing most negative flux are dependent on $\hbar$ (and $m$ and $T$). Numerical work indicates that the most negative eigenvalue increases with $a$. It is still negative for finite $a$, indicating that backflow will still be present for more realistic measurements. The lowest eigenvalue appears to go to zero for $a \rightarrow \infty$. This indicates that all the negative eigenvalues go to zero (or become positive)
in the naive classical limit $\hbar \rightarrow 0 $.
(This is in contrast to the more realistic classical limit of a particle coupled to an environment, where it can be shown that the backflow effect does indeed vanish exactly after a finite time \cite{Ye1}). By contrast the positive eigenvalues are not significantly affected.

\section{Backflow and Measurement Models}

In this section we relate the above results on backflow to measurements. This is partly to begin to address the practical question of how backflow is measured, but also to get some insight into the negativity of the flux.
A discussion of the possible measurement of backflow was given by Bracken and Melloy \cite{BrMe} and some earlier discussions of measurement of the current may be found in Refs.\cite{Ber,GGT,AhVa,DDGZ,MBH}.

\subsection{Explicit Measurement of Backflow}

Our first observation concerning the possible measurement of backflow
is that the flux Eq.(\ref{flux}) is defined as the difference between two probabilities,
therefore the flux can be measured by measuring these two probabilities.
This requires {\it two} ensembles, each prepared in the initial state $ | \psi \rangle$.
On one ensemble, measurements are made to determine if the particle is in $x>0$
at time $t_1$, hence determine $\langle P(t_1) \rangle$.
On the second ensemble, the same measurements are performed at time $t_2$,
which thus determines $\langle P(t_2) \rangle$. From the two results the flux can determined. 

This is perhaps the most direct way of measuring backflow and could in principle be done using Bose-Einstein condensates. Briefly, for weak interactions a Bose-Einstein condensate corresponds to a whole ensemble of non-relativistic particles, so measurements of the above probabilities could be determined by a single measurement on the condensate \cite{MugaBE}. Backflow could be investigated if it is possible to prepare the system in a state of positive momentum. This will be explored in more detailed in a future publication.

Note that the above is not the same as performing sequential measurements of position
on the {\it same} ensemble. We will come to these sorts of measurements below.


\subsection{A Simple Measurement Model for Arrival Time}

A different way of gaining insight into the properties and measurement of the flux is to consider simple models for measuring the arrival time. This is because such models, if properly constructed, yield a non-negative probability which will, however, be approximately the same as the flux in some limit, since the flux is the correct semiclassical probability for the arrival time
\cite{time,HaYe1,HaYe2,YDHH}. Hence by comparing the (always non-negative) probability arising in such models with the (sometimes negative) flux we may be able to see the origin of the negativity and also gain some insight into ways in which the current can be measured.

The simplest model for measurement of the arrival time involves simply measuring
to see if the particle is in $x<0$ at time $t_1$ and then in $x>0$ at time $t_2$.
This probability is given by
\beq
p(t_1, t_2) = \langle \psi | \bar P (t_1) P(t_2) \bar P (t_1) | \psi \rangle
\label{prob0}
\eeq
which is clearly positive. It gives a simple notion of arriving at the origin
during the time interval $[t_1,t_2]$ (but ignoring issues about multiple crossings).
Using the flux operator Eq.(\ref{fluxop}) this may be rewritten
\bea
p(t_1, t_2) &=& \langle \psi | \bar P (t_1) ( P(t_2) - P(t_1) ) \bar P (t_1) | \psi \rangle \nonumber \\
&=& \int_{t_1}^{t_2} dt \ \langle \psi | \bar P(t_1) \hat J(t) \bar P (t_1) | \psi \rangle
\label{prob}
\eea
This coincides with the flux Eq.(\ref{flux2}) except for the projection operators
onto $x<0$ at $t_1$. Since Eq.(\ref{prob}) is positive, this means that the negativity of
the flux comes entirely from the part of the state which is already in $x>0$
at the initial time $t_1$.

For a wave packet which either cleanly crosses or does not cross the origin, during the time interval $[t_1,t_2]$,
the probability Eq.(\ref{prob}) will, to a good approximation, be equal to the flux, which will be positive (or zero). However, for states with backflow, the flux is negative but $p(t_1,t_2)$ is non-negative, so there will be a substantial difference between them. The interesting question is then to see how the negativity of the flux leaves its signature in the non-negative arrival time probability. To see this we need a more elaborate model.

\subsection{A Complex Potential Model for Arrival Time}


Many more elaborate and realistic models for the measurement of the arrival time (involving model detectors, for example) naturally lead to an arrival time probability defined with a complex potential. This is described in detail in many places \cite{HaYe1,HaYe3,complex,Ech,Hal3}. These models typically yield an arrival time probability distribution which is closely related to the current and from which the current may be extracted, even when negative, thereby leading to a possible measurement of backflow.

A typical model is something like the following. We again consider an initial wave packet starting in $x<0$ with positive momentum and seek the arrival time probability distribution $\Pi (\tau)d \tau $ for crossing the origin between $\tau$ and $\tau + d \tau$. We consider a complex absorbing potential of step function form in $x>0$
so the Hamiltonian is $H_0 - i V_0 \theta (\hat x) $, where $H_0$ is the free Hamiltonian. We define the survival probability
$N(\tau)$ to be the norm of the state at time $\tau$ after evolution with this complex Hamiltonian.
The arrival time distribution
is then given by
\bea
\Pi (\tau) &=& - \frac {d N} {d \tau}
\nonumber \\
&=& 2  V_0 \langle \psi | e^{ \left(  i H_0   - V_0 \theta (\hat x)   \right) \tau} \theta (  \hat x ) e^{\left( - i H_0   - V_0 \theta (\hat x)    \right) \tau } | \psi \rangle
\label{A.1}
\eea
We seek a simple form for this expression which exposes its dependence on the current operator and thus gives some idea as to how it will be affected when backflow is present.
Differentiating with respect to $\tau$, we get
\beq
\frac {d \Pi } {d \tau} = - 2V_0 \Pi  + 2  V_0 \langle \psi | e^{ \left(  i H_0   - V_0 \theta (\hat x)    \right) \tau } \hat J e^{\left( - i H_0   - V_0 \theta (\hat x)    \right) \tau} | \psi \rangle
\label{A.2}
\eeq
where $\hat J$ is the current operator Eq.(\ref{curop}).
Eq.(\ref{A.2}) is a differential equation for $ \Pi (\tau) $ which is easily solved to yield
\beq
\Pi (\tau) = 2  V_0 \int_{-\infty}^\tau dt \ e^{ - 2 V_0 (\tau - t) }
\ \langle \psi | e^{ \left(  i H_0   - V_0 \theta (\hat x)   \right)t } \hat J e^{\left( - i H_0  - V_0 \theta (\hat x)    \right)t } | \psi \rangle
\label{A.4}
\eeq
where we have assumed that $\Pi (\tau) \rightarrow 0 $ as $ \tau \rightarrow - \infty$.
Eq.(\ref{A.4}) is the exact expression for $\Pi (\tau)$ and displays the dependence on the current operator $\hat J$.
It is positive by construction, even though $\hat J$ is not a positive operator.
The probability for crossing during the time interval $[t_1,t_2]$ then is
\beq
p(t_1, t_2) = \int_{t_1}^{t_2} dt \ \Pi (t)
\eeq
This is the analogue of Eq.(\ref{prob}).

It is not easy to see how the presence of backflow states in Eq.(\ref{A.4}) may register in the
probability distribution $\Pi (\tau)$. The expression is, however, simpler in
the usual weak measurement approximation (small $V_0$), which involves neglecting the complex potential terms in the bracket expression, yielding
\beq
\Pi (\tau) \approx 2  V_0 \int_{-\infty}^\tau dt \ e^{ - 2 V_0 (\tau - t) }
\ \langle \psi_t | \hat J  | \psi_t \rangle
\label{A.5}
\eeq
where $ | \psi_t \rangle = e^{-iH_0 t} | \psi \rangle $. This is the expected semiclassical result \cite{All,DEHM,HSM,HaYe1} (although note that the new derivation given here is considerably shorter than those given elsewhere!).

Note that Eq.(\ref{A.5}) is not necessarily positive, due to the negativity of the current in certain states. The positivity may have been
lost in going from Eq.(\ref{A.4}) to Eq.(\ref{A.5}) because we took the limit $V_0 \rightarrow 0$
in the bracket expression but not in the rest of the expression. However, this should not matter for sufficiently small $V_0$, and we will assume that Eq.(\ref{A.5}) is positive.

The quantity $\Pi (\tau)$ corresponds to the arrival time distribution measured by a realistic measurement so can in principle be determined experimentally. The current can then be extracted from Eq.(\ref{A.5}) by deconvolution \cite{DEHM,HSM} or by taking a derivative, via Eq.(\ref{A.2}) (with the limit 
of small $V_0$ taken in the current expression). This therefore gives a method of measuring the current and hence the flux, and checking for backflow.

Eq.(\ref{A.5}) has the form of the current smeared over a range of time. This general form has also been observed in other models for the measurement of arrival time (see for example Ref.\cite{YDHH}). What this means is that a region of negative current can cancel out a region of positive current in the measured probability $\Pi (\tau)$.
This may be interpreted as meaning that backflow produces a time delay between the arrival of the wave packet at the origin and its registration in a measuring device. (Ideas along these lines were explored in Ref.\cite{Muga}).

In Fig.14 we plot the measurement probability Eq.(\ref{A.5}) for two values of $V_0$ and also the original numerically computed current, to see how the time-smearing affects the backflow. We see that positive regions of the current are not qualitatively changed very much, in keeping with semiclassical expectations, but negative regions of the current become positive as a result of the smearing, as they must, since the measured probability is positive.

It is also striking that the discontinuous jumps in the current from positive to negative at $t= \pm 1$ arise as discontinuous changes in the derivative of the time-smeared current. We speculate that such discontinuities may be signatures (in the measured probabilities) of backflow in the underlying current.

\begin{figure}[htbp]
   \centering
   \includegraphics[width=5in]{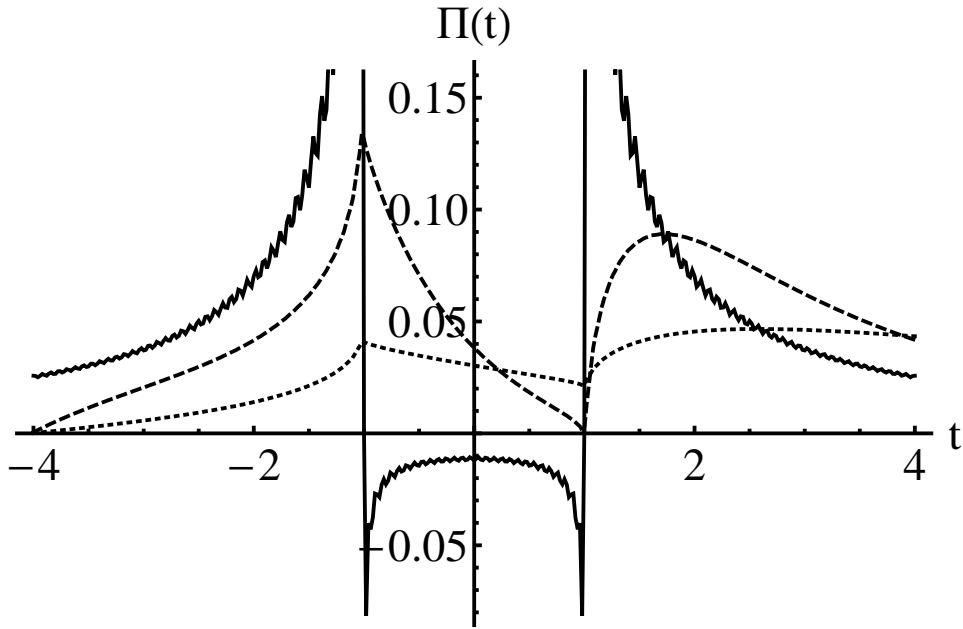}
   \caption{A plot of the current (solid line) and time-smeared current Eq.(\ref{A.5}) for $V_0=0.5$ (dashed line)
   and $V_0 = 0.1 $ (dotted line).}
   \label{Fig14}
\end{figure}



\section{Summary and Conclusions}

The purpose of this paper was to explore and illustrate the backflow effect in a variety of different ways. After setting up the problem in Section 2, in Section 3, we computed the current and the flux for states consisting of superpositions of gaussians. These states are important since they are experimentally realizable. Backflow is easily obtained with these states but the maximum amount of flux is very small, only about 16 percent of the maximum possible.

In Section 4, we looked for analytic expressions for states matching as closely as possible the numerically computed states giving maximal backflow, computed by Penz et al \cite{Penz}. We presented two candidate analytic expressions and computed the current at arbitrary times analytically. The plot of the current in each case had reasonably good agreement with the numerical solution, except at the end points $t = \pm 1$ of the backflow region. We computed the most negative flux of these states.
In one case, the total flux is about 70 percent of the numerically computed maximum backflow, significantly better than any previous analytic expression for a backflow state.
For this most negative flux state, we plotted the probability of remaining in $x<0$ against time in Fig.(\ref{figptapprox}). This gives a particularly striking illustration of the backflow phenomenon, showing a distinct period of increase in probability.

Note that although the backflow obtained in these analytic guesses is significant, which is what we aimed to achieve,
it is not in fact that close to the maximum backflow, despite the fact that our analytic guesses for the momentum space wave function appeared to be very close. What is perhaps relevant here is that our analytic wave functions failed to match the singularity structure of the current at $t= \pm 1$. We deduce from this that the singularity structure of the current is somehow important in obtaining the maximum backflow states. This issue will be addressed in future publications.

In Section 5, we discussed the classical limit of backflow. The most interesting aspect of this is the issue, first noted by Bracken and Melloy, that the eigenvalues of the flux operator are independent of $\hbar$. This appears to mean that there is a genuine quantum phenomenon, negative flux, which is independent of $\hbar$ and which does not appear to go away in the naive classical limit $\hbar \rightarrow 0 $. We showed that this situation starts to appear more physically sensible when the projectors used in the definition of the flux operator are replaced by quasiprojectors, which includes a physical parameter characterizing the imprecision of real measurements. The eigenvalues then do depend on $\hbar$ and the most negative eigenvalue becomes less negative as the quasiprojectors become more smeared. Furthermore, there is evidence that all the negative eigenvalues become zero or positive as $\hbar \rightarrow 0 $, restoring the naive classical limit.
However, there are clearly more issues to explore around this question.

In Section 6 we discussed measurement models that exposed certain aspects of backflow. Eq.(\ref{prob}) establishes that backflow arises from the part of the state which is already in $x>0$. The complex potential model of Section 6(C) corresponds to a number of reasonable realistic measurement models. The current can be obtained from the measured probability by deconvolution, and from this result the negative current could in principle be obtained.
(Along the way, we also discovered a very concise derivation of the arrival time formula with a complex potential, Eq.(\ref{A.5})).


The features of backflow elucidated here may be of value in designing experiments to test backflow. These and related ideas with be explored elsewhere.

Added note: After completion of this work we became aware of interesting related work involving the backflow effect for angular momentum \cite{strange}.

\section{Acknowledgements}

We are very grateful to Gonzalo Muga for many useful conversations about the topic of this paper.
We would also like to thank Tony Bracken, Markus Penz and Gebhard Gruebl for useful comments on an earlier draft of this article. JMY was supported by the John Templeton Foundation.

\bibliography{apssamp}


\end{document}